\documentclass[sn-apa]{sn-jnl}

\jyear{2021}%

\theoremstyle{thmstyleone}%
\theoremstyle{thmstyletwo}%

\theoremstyle{thmstylethree}%

\newcommand\cites[1]{\citeauthor{#1}'s\ (\citeyear{#1})}

\begin{document}

\title[A Network Approach to Consumption]{A Network Approach to Consumption}


\author*[1]{\fnm{Jan} \sur{Schulz}}\email{jan.schulz@uni-bamberg.de}

\author[2]{\fnm{Daniel M.} \sur{Mayerhoffer}}\email{daniel.mayerhoffer@uni-bamberg.de}

\affil*[1]{\orgdiv{Economics Department}, \orgname{University of Bamberg}, \orgaddress{\street{Feldkirchenstr. 21}, \city{Bamberg}, \postcode{96052}, \state{Bavaria}, \country{Germany}, \href{jan.schulz@uni-bamberg.de}{mailto:jan.schulz@uni-bamberg.de},
Phone number: [+49] (951) 863-2729, ORCID-ID: 0000-0001-7745-3997}}

\affil[2]{\orgdiv{Institute for Political Science}, \orgname{University of Bamberg}, \orgaddress{\street{Feldkirchenstr. 21}, \city{Bamberg}, \postcode{96052}, \state{Bavaria}, \country{Germany}, ORCID-ID: 0000-0001-8841-407X}}

\abstract{The nexus between debt and inequality has attracted considerable scholarly attention in the wake of the global financial crisis. One prominent candidate to explain the striking co-evolution of income inequality and private debt in this period has been the theory of upward-looking consumption externalities leading to expenditure cascades. We propose a parsimonious model of upward-looking consumption at the micro level mediated by perception networks with empirically plausible topologies. This allows us to make sense of the ambiguous empirical literature on the relevance of this channel. Up to our knowledge, our approach is the first to make the reference group to which conspicuous consumption relates explicit. Our model, based purely on \emph{current} income, replicates the major stylised facts regarding micro consumption behaviour and is thus observationally equivalent to the workhorse permanent income hypothesis, without facing its dual problem of `excess smoothness' and `excess sensitivity'. We also demonstrate that the network topology and segregation has a significant effect on consumption patterns which has so far been neglected.}




\maketitle

\section{Introduction}\label{sec1}
When the potentially destabilising effects of both economic inequality and private debt have become apparent in the wake of the global financial crisis, the nexus between the two has also increasingly attracted scholarly interest. One prominent candidate to explain the striking co-evolution of inequality and private debt in this period has been the theory of upward-looking consumption externalities \citep{vantreeck2014}. Thereby, the increases in the top relative incomes are assumed to have triggered debt-financed conspicuous consumption of the poorer segment of the population. While several microeconomic studies testify to the relevance of such conspicuous consumption \citep{heffetz2011,bertrand2016, bricker2020}, the macroeconomic evidence on this channel has been somewhat limited and ambiguous \citep{wildauer2018, bofinger2019}. We propose a parsimonious model of upward-looking consumption mediated by perception networks with empirically plausible topologies to reconcile these two contradictory findings. Up to our knowledge, our approach is the first to make the reference group to which conspicuous consumption relates explicit.\footnote{The foundational contribution by \cite{Frank2014} only builds on classifications according to income deciles and does not include a discussion of inner-decile links that \cite{schulz2022} find to be of crucial importance for perceptions.} As predicted by the (upward-looking) relative income hypothesis, aggregate savings unanimously decrease in inequality. However,  endogenously evolving perception networks featuring homophily might mitigate this effect because an increase in inequality also increases segregation within the network. For high degrees of homophily, the effect vanishes, potentially explaining the mismatch between micro-behaviour and macro-aggregates. 

Our model is an attempt to answer the challenge posed by \cite{nelson2010} that its Schumpeterian heritage lets evolutionary economics exhibit a supply-side bias and neglect demand-side factors. In particular, we build on two central themes of evolutionary economics: The relevance of imitation and of meso-level structures for macroeconomic outcomes. Following \cite{veblen2001}, individuals engage in conspicuous consumption activity, triggering imitative and emulative behaviour by individuals with lower social status \citep{nelson2010}. Recent research indeed suggests conspicuous consumption itself to be likely of evolutionary origin, though emerging not from sexual selection but rather cultural evolution within contemporary market societies \citep{koliofotis2021}. This culturally evolutionist perspective implies that conspicuous consumption is a socially learned activity acquired through observation and imitation \citep{witt2001, koliofotis2021}. As a direct corollary, individuals must be able to \emph{observe} others' consumption practices in order to emulate them, which we formalise by a plausible perception network following \cite{schulz2022}.
This complex interaction at the meso-level thus mediates and shapes the effect of micro-level behaviour (i.e., individual consumption) on macro-level outcomes (i.e, expenditure distributions and aggregate consumption levels) \citep{foster2021}.

The argument that intersubjective comparisons can be powerful personal motivators traces back to at least Adam \cites{smith1991} \emph{Wealth of Nations}.\footnote{Cf. \cites{sen1984} foundational discussion about absolute and relative dimensions of poverty.} He famously argued that even something as simple as wearing a linen shirt might be a ``necessary of life in England'' \citep[vol. ii, p. 352]{smith1991} at the time of his writing to signal that the wearer conforms to custom. Status signalling has been a central tenet, especially of heterodox theories of consumption behaviour ever since \citep{veblen2001,Duesenberry1949,witt2001,Frank2014}. However, while most those theories focus on luxury goods, \cite{smith1991} reminds us that status goods are essentially defined by custom, and that their signalling is not an inherent feature of them.  Therefore, status consumption is not confined to a specific social class that can afford expensive luxury goods but is a human tendency that might apply to all individuals across the whole income distribution. Yet, not all consumption is status consumption: \cite{keynes1972} divided goods into two types: Non-positional goods relating to material needs and status goods relating to needs ``which are relative in the sense that we feel them only if their satisfaction lifts us above, makes us superior to, our fellows'' \citep[p. 326]{keynes1972}. In the terminology of an evolutionary theory of consumption, this taxonomy would roughly correspond to basic, innate needs `needs' and socially constructed `wants' \citep{witt2001, rengs2019}. Psychological evidence indeed corroborates the conjecture that status consumption is upward-looking \citep{frank1985} and that both absolute and relative motives matter \citep{alpizar2005}.

Our model includes both the Smithian and Keynesian insight and lets \emph{all} consumers have a propensity for \emph{upward-looking} status consumption, with the whole population consuming a basket of both status and non-status goods. This parsimonious consumption rule based only on \emph{current} income replicates all the major stylised facts regarding micro consumption behaviour. It is thus at least observationally equivalent to the workhorse permanent income hypothesis, without facing its problems of `excess smoothness' and `excess sensitivity' \citep{meghir2004}.

Despite their emphasis on signalling and perception, formal theories of conspicuous consumption seldomly explicate to which others the consumers exactly aspire to catch up to. Typically, only global or class average consumption levels are considered \citep{alpizar2005, rengs2019, alvarez2011, petach2021}, meaning that local structures are not part of these explanations. Contrary to that, the sociological literature indicates that empirical social networks are far from random and consequently that human interaction is strongly structured. The major contribution of this paper is to show that the impact of this structure, most notably homophily or the tendency to link to individuals that are similar to oneself \citep{lazarsfeld1954}, is non-negligible and might help explain some of the recent empirical puzzles. The most notable of these puzzles is the mismatch between significant findings in the applied micro literature and the ambiguous findings from more macroeconomically oriented approaches on positional concerns. We propose a new channel, where income inequality not only incentivises conspicuous consumption but also increases segregation due to income homophily, thereby decreasing \emph{perceived} inequality and mitigating the initial effect of \emph{actual} inequality. \cite{bertrand2016} document the relevance of this channel for residential segregation, and we hope to inspire further empirical studies with our model exercise, especially in the direction of occupational or educational segregation in relation to consumption decisions.

The remainder of this paper is organised as follows: Section~2 situates our model within the pertinent literature and shows how both sociological and economic theory inform our modelling choices. Section~3 introduces our formal model both regarding consumption behaviour and the underlying social network. Section~4 presents our simulation results. Within the section, we demonstrate both that the emergent expenditure distributions are consistent with the relevant stylised facts and that the endogenous adjustment of the social networks can generate both nonlinearities and a large variety of point elasticities, underlining the potential relevance of the segregation channel. The final section~5 concludes and discusses possible avenues for further research, particularly possible empirical applications and modelling extensions to include our partial model of consumption decisions within a full-fledged macro framework. 

\section{Related Literature}\label{sec:RelatedLiterature}
The advent of \cites{friedman1957} permanent income hypothesis was a paradigm change in the theory of consumption behaviour and shifted the focus from explaining consumption based on current income to explaining it with the concept of permanent income composed of unobservable stochastic income shocks. The subsequent literature has focused mainly on refining Friedman's initial hypothesis by including expectations, precautionary savings motives or liquidity constraints to bring it closer to the data \citep[for surveys]{palley2010, meghir2004}. Our focus is different. We abstract from any intertemporal expectation formation or additional liquidity constraints and show how the emergent expenditure distributions relating to \emph{current} income only can give rise to the relevant empirical stylised facts of consumption behaviour purely from human interaction within plausible social networks. Nonetheless, expectations regarding future income streams are potentially relevant for consumption behaviour, and our model merely demonstrates that the empirical findings on consumption expenditures both on the micro and on the macro level do not necessarily imply intertemporal optimisation over (largely unobservable) future income streams, as is typically argued in the literature \citep{battistin2009}. In contrast to these demanding assumptions on (essentially unbiased or model-consistent) individual expectations, we make the perceptions of each agent fully explicit without any kind of expectation formation on future income streams to  Thereby, we bring two strands of literature together to make sense of a third one: Namely, we combine the rich extant literature on empirical social networks with the empirical evidence on upward-looking consumption externalities to provide a new explanation for the ambiguous empirical findings on the aggregate savings-inequality nexus.

The topology of empirical social networks exhibits salient and universal features that can serve as stylised facts to validate artificial graphs. These include the small-world property which holds across many different domains \citep{galaskiewicz2007, newman2001, uzzi2005, weeden2020}, homophily in tie-formation, especially for economic class \citep{boucher2015, cepic2020, huckfeldt1983, malacarne2017, mayer2008} and their sparsity, especially when focusing on close ties \citep{Maccaron2016, degiorgi2020}. \cite{schulz2022} present a model of homophilic tie formation that can generate these stylised facts of empirical graph topologies as well as being able to replicate all the relevant findings on positional self-assessment and perceptions of inequality. Since our argument builds on status consumption, which rests on the individuals' perception of consumption, we adopt this modelling framework as an algorithm for plausible social networks.

One can understand this as individuals trying to maximise their social capital \citep{coleman1988,annen2003} by picking those others as social contacts who are like them in certain respects because this minimises the transaction costs of social interaction \citep{akerlof1997}. Such transaction costs may simply be travel time if the social contacts do not live nearby, effort necessary to understand each other's cultural and social background or their interests. However, as factors like residential area \citep{hu2022, harting2020}, education \citep{smith2014,leo2016}, ethnicity \citep{chandra2000}, lifestyle \citep{vrtanen2007}, or even health \citep{krieger1992} correlate with income, the latter constitutes a decent proxy for the actual homophilic behavioural patterns. Moreover, there is also homophily in income surfacing in, e.g. mobile phone communication \citep{xu2021,leo2016,fixman2016}. Therefore, while people may not willingly choose their social contacts based on income proximity, their actual choices amount to social contacts as if selected homophilic in income.

The importance of interpersonal comparisons in consumption and income is well established for the microeconomic level. Among others, \cite{bertrand2016}, \cite{bricker2020} and \cite{heffetz2011} find significant effects of interpersonal comparisons for individual consumption in the US, \cite{jinkins2016} for the US and China, \cite{quintana2016} for the UK, \cite{drechsel2014} for Germany and \cite{alpizar2005} for Costa Rica. However, the choice of reference groups within these studies often lacks granularity, underlining the need for more plausible models of interaction in networks. The empirical psychological literature reviewed in chapter 2 of \cite{frank1985} only establishes that conspicuous consumption is generally \emph{upward-looking}. Given upward-looking consumption externalities, uneven income gains that are skewed to the top can trigger ``expenditure cascades'' if the poorer try to emulate the consumption behaviour of the richer \citep{Frank2014}. Indeed, several studies for the US also document the relevance of inequality for private debt-buildup of the poor to enable consumption increases, also in line with the relative income hypothesis \citep{agarwal2020, carr2015, christen2005, schmid2013}.

Given the apparently unambiguous results from more micro-level studies, it seems surprising that this view has not manifested itself in a consensus on the macro-level association of inequality and (private) savings. While \cite{koo2016} report that aggregate saving rates increase with income inequality due to the rich saving more, \cite{klein2015}, \cite{perugini2016}, \cite{behringer2018}, \cite{behringer2019} and \cite{petach2021} find evidence for a negative relationship between savings and inequality, in line with the expenditure cascades hypothesis. Most studies fail to find a significant effect in either direction, though \citep{bordo2012, cuaresma2018, gu2014, gu2015, leigh2009, schmidt2000, wildauer2018}. \cite{cuaresma2018}, \cite{gu2014} and \cite{gu2015} highlight the potential relevance of unobservable intermittent country characteristics shaping the relationship between savings decisions and inequality, which might in part explain why \cite{bofinger2019} find a strongly non-monotonic relationship between both.\footnote{\cite{bofinger2019} include country- and time-fixed effects into their estimation. These might, however, be insufficient to capture time-variant country idiosyncrasies, as our endogenously evolving social network might suggest.} Furthermore, \cite{wildauer2018} highlight the time-scale of adjustments and show that with the proper controls and examining a long-run relationship, the effect of inequality on savings seizes to be significant.

We conclude that a proper model of expenditure cascades has to make sense of the variability of effect size as a function of the considered time scale. Moreover, such a model should give a plausible account of the origin of cross-country heterogeneity. To demonstrate the relevance of network structures for these phenomena, we use an otherwise purely upward-looking consumption rule inspired by \cite{Frank2014} that should in and of itself give rise to an unambiguously negative relation between savings and inequality but can generate a large variety of effect sizes as a function of network topology. Finally, we require our model to be consistent with the major stylised facts on empirical expenditure distributions for validation at the micro-level.


The empirical literature has identified four stylised facts regarding consumption expenditures and their relation to (current) incomes: (i) individual average propensities to consume (APCs) tend to decrease in (current) income \citep{dynan2004,fagiolo2019}; (ii) population-level APCs remain roughly constant with respect to changes in total income \citep{kuznets1942}; (iii) the distribution of consumption expenditures is more homogeneous than the distribution of current income \citep{krueger2006,jappelli2010} and (iv) the distribution of consumption expenditures is (at least as a first-order approximation) well fit by a log-normal distribution \citep{battistin2009, brzozowski2010, chakrabarti2018, hohnisch2002, fagiolo2010, ghosh2011, toda2017}, while the distribution of current income is not \citep{dragulescu2001, silva2004, tao2019}. The last stylised fact constitutes evidence for the log-normality of expenditure distributions since several of those studies also find significant deviations from the log-normal benchmark, particularly for their upper tails \citep{chakrabarti2018,fagiolo2010,ghosh2011,toda2017}. In light of these findings, we also discuss the relevant parameter range of our model for which log-normality holds, namely, for intermediate ranges of social consumption to make sense of possible deviations from the log-normal benchmark that can also emerge within our model. 

\section{Model}\label{sec:Model}
This section gives a content-oriented presentation; we provide a technical description following the ODD protocol, the NetLogo implementation of the simulation model on \href{https://github.com/mayerhoffer/Inequality-Perception}{https://github.com/mayerhoffer/Inequality-Perception}. The model consists of two distinct phases running in sequential order, where phase one builds the environment and determines available information for the consumption procedure in the second phase:
\begin{enumerate}
	\item Network generation
	\begin{itemize}
		\item Agent initialisation and income allocation
		\item Homophilic linkage
	\end{itemize}
	\item Individual perception and consumption
\end{enumerate}
\subsection{Network Generation}
There are 1000 agents in the model; each agent draws their income from an exponential distribution with a mean of $\lambda = 1$. Such a distribution normalises the empirical observed (pre-tax or market) income distributions in various industrialised countries for the vast majority of individuals \citep{dragulescu2001,silva2004,dos2017,tao2019}.\footnote{Interestingly, \cite{dos2017} shows how an exponential wage distribution emerges from social comparisons, much like the consumption externality we model in our consumption rule. It, therefore, appears promising to bring these two modelling approaches together.} Thus, one can understand the model population as constituting a representative sample of empirical populations of these countries. The upper tail of $1$ to $5$ \% of the income distributions empirically follows a Pareto law \citep{silva2004}. We deliberately choose to exclude this small minority from our model since their population size would induce another degree of freedom in our model, and we want to demonstrate that segregation is indeed endogenous and not driven by differences in the actual income regime. Hence, we also use log-normal distributions for comparison between different inequality levels. While such log-normal distributions do not represent empirical findings, they allow for varying the inequality level through their dispersion parameter $\sigma$ without changing the overall distribution shape. Hence, the present model employs the log-normal distribution solely to evaluate the relationship between actual income inequality, individual perceptions and consumption behaviour. Agents store their true income decile for evaluation purposes, too.

Each agent draws five other agents to link to. Like for real-world networks, links are therefore created \emph{by agents}, not imposed on them. The number of five-link choices represents the closest layer of intense contacts identified in empirical studies \citep{zhou2005, hamilton2007, Maccaron2016}\footnote{For a recent review on the large literature on 'Dunbar's number`, cf. the first section of \cite{Maccaron2016}.}. While this closest layer consists of differently related people for different individuals \citep{karlsson2005}, what those individuals do, \cite{wilcox2013} empirically identify this layer as the relevant reference group for social consumption. Moreover, we have carried out sensitivity analyses and found a larger number of links to have little impact on overall simulation results. Agent $j$'s weight in agent $i$'s draw is denoted by $\Omega_{ij}$ and determined as follows:
\begin{align}
		\Omega_{ij}&=\frac{1}{\exp[\rho~\mid Y_{j} - Y_{i}\mid]}\label{networkformation}
\end{align}
The relative weights in the draws are a function of the homophily strength and the respective income levels: $Y$ denotes the income of an agent, and $\rho \in \mathbb{R}^+_0$ denotes the homophily strength in income selection, externally set, and identical for all agents. $\rho = 0$ represents a random graph, and for an increasing positive value of $\rho$, an agent becomes ever more likely to pick link-neighbours with incomes closer to their own. The link function's exponential character ensures that those with large income differences become unlikely picks even at low homophily strengths. For an extensive analytical discussion of this linkage behaviour, see \cite{schulz2022}. While our linkage function represents a tractable reduced-form representation of the empirically well-established homophily, \cite{falk2004} demonstrate that endogenous reference groups can emerge from the social comparisons of agents aiming to self-improve and self-enhance. Similarly, our weighting function can be interpreted as agents selecting contacts as if minimising the income distance that they observe as a noisy signal. As discussed in Section~\ref{sec:RelatedLiterature}, this assumption does not require individuals to consciously form their social ties based on income proximity. Instead, they may exhibit homophily in e.g. education level, ethnicity, or lifestyle; but thereby the individuals act as if there was explicit income homophily because the aforementioned factors correlate with income. As a consequence, segregation increases with income inequality, as is well documented in the literal sense of residential segregation \citep{wheeler2008, watson2009, reardon2011, chen2012, toth2021}. However, linkages in our sense extend beyond this literal geographic sense and essentially imply that consumption decisions are observable, be it due to common workplaces, family ties or any other form of linkage.

Formally, we assume that the utility of agent $i$ connecting to agent $j$ follows an additively separable utility function with agents receiving disutility proportionally to the absolute income distance to the income of agent $j$ and a random error term, i.e.,
\begin{align}
	U_{ij} &= -\rho \mid  Y_j - Y_j \mid  + \epsilon_{ij}.
\end{align}
When the distribution of $\epsilon_{ij}$ is identically and independently distributed according to an extreme value distribution, we can express the choice probability of agent $i$ as
\begin{align}
	p_{ij} &= \frac{\exp[-\rho \cdot \mid Y_{j} - Y_{i} \mid ]}{\sum_{k \in M \setminus i}\exp[-\rho \cdot \mid Y_{k} - Y_{i} \mid },\label{discretechoice}
\end{align}
with $M \setminus i$ as the set of all agents except $i$ with size $N - 1$ which is equivalent to the weights in eq.~\eqref{networkformation} translated into probabilities \citep{hoffman1988}. Our weighting function is thus an application of the discrete choice approach developed and popularised by \cite{manski1981}. The random utility model appears to us to be particularly appropriate here, since income is a rather salient characteristic to determine a good `fit' within a social network but of course might depend on other characteristics that are not directly observable and thus modelled to be stochastic.\footnote{Note, however, that this choice rule implies the axiom of Independence of Irrelevant Alternatives (IIA) to hold for all alternatives and was in fact explicitly designed to do so \citep{manski1981}, i.e., the probability of choosing between $j$ and $k$ to be independent of the probability of choosing $l$ \citep{luce1977}. IIA might be considered a good first-order approximation to model homophilic choice, but especially within social networks, knowing one agent $j$ might indeed increase the likelihood of knowing another agent $l$ that is acquainted with $j$. It might thus prove interesting to extend and generalise the above rule for correlated choice to examine the effects on the network topology in further research.} The homophily parameter $\rho \in (0,\infty)$ is then simply the intensity of choice parameter with $\rho \to \infty$ implying that $i$ chooses agent $j$ to link to with certainty who has minimal income distance, i.e., $p_{ij} \to 1$.  In this sense, the weighting function in eq.~\eqref{networkformation} is plausibly microfounded and can now be considered the workhorse choice rule in behavioural macroeconomics \citep{franke2017}. \cite{franke2017} also survey evidence from a several lab experiments in different macroeconomic contexts that discrete choice is indeed consistent with the data, while \cite{anufriev2012} and \cite{anufriev2018} provide laboratory evidence for the discrete choice approach for financial markets.

\subsection{Individual Perception and Consumption}
Generally, positional concerns can be modelled both within an intertemporal maximisation framework with explicit expectation formation \citep{drechsel2014} or building merely on current income without any expectations relating to future income streams \citep{Frank2014}. We choose the latter option for two reasons: Firstly, we aim to keep assumptions on individual behaviour as minimal as possible to preserve maximal generalisability. The workhorse consumption model built on Euler equations as famously introduced by \cite{hall1978} builds on rational expectations that are very demanding as an assumption. Empirically, \cite{pesaran2006} find little evidence for the rational expectations hypothesis. Theoretically, \cite{hendry2014} show that a violation of the usual regularity conditions in stochastic processes like non-stationarity or structural breaks implies that rational forecasts conditioned on past performance are suboptimal, essentially since the application of the law of iterated expectations is hindered. Thus, it seems highly unlikely that the rational expectations benchmark indeed holds in the real world, especially if income processes are very noisy. Secondly, the dual problem of `excess smoothness' and `excess sensitivity' for calibrated consumption models of permanent income hints at the shortcomings of the permanent income hypothesis \citep{meghir2004}, that is, consumption reacting too strongly (weakly) to (un)anticipated income shocks.\footnote{The commonly employed habit persistence extension to remedy these problems yields coefficients of habit formation that are typically deemed much too high to be plausible \cite[p. 887 -- 892, and the references therein, for a more detailed discussion]{pesaran2015}. Coincidentally, \cite{binder2001} demonstrate that habit persistence in conjunction with social comparisons might be able to solve both problems simultaneously.} By contrast, our model is based explicitly on current income (and intersubjective comparisons) and does therefore not run into the problem that consumption is tracking current incomes too closely. We also do not need to impose an unrealistic degree of heterogeneity or additional assumptions to match the empirically observed heterogeneity in consumption propensities \citep{jappelli2014}.


In particular, the only heterogeneity we impose is the (pre-validated) income distribution. Agents consume maximising utility $U(\cdot)$ with Cobb-Douglas preferences and derive (linear) disutility the stronger their consumption in isolation given by $\tilde{C}(i)$ falls short of the highest consumption level $C(j\mid i)$ they observe within their ego network with intensity parameter $c > 0$. With this assumption, we follow the tradition of additive comparison utility in contrast to ratio comparison utility, where others' consumption enters as a ratio to own isolated consumption \citep{alpizar2005}.\footnote{A prominent example of the additive approach is \cite{akerlof1997}, while one example of ratio comparisons can be found in \cite{carroll1997}. Cf. also the literature review in \cite{alpizar2005}. The qualitative intuition should be equivalent in both cases, though.} Apart from that the utility function is rather standard with agents deriving utility both from consumption with elasticity parameter $w$ and saving with elasticity parameter $(1-w)$ to capture intertemporal motives in the most parsimonious fashion. The choice variable is the average propensity to consume $\gamma$, i.e., the fraction of current income $Y$ an agent consumes. The utility function of agent $i$ thus reads
\begin{align}
	U(\gamma; Y(i),C(j\mid i), \tilde{C}(i), c,w) &= (\gamma_i Y(i) - c\cdot(C(j\mid i) - \tilde{C}(i) ))^w \cdot \nonumber \\ &~~~((1-\gamma)Y(i))^{1-w}\label{gen_utility}
\end{align}

The utility function for consumption in isolation, $\tilde{U}$ with choice variable $\tilde{\gamma_i}$ is given by eq.~\eqref{gen_utility} for $c = 0$ and thus reads
\begin{align}
	\tilde{U}_i(\tilde{\gamma_i};w,Y(i)) &= (\tilde{\gamma_i}Y)^{w}(Y_i\cdot(1-\tilde{\gamma_i}))^{1-w}.
\end{align}
The FOC for optimal $\tilde{\gamma}$ reads
\begin{align}
	\tilde{\gamma}_i &= w,
\end{align}
and thus optimal consumption without social interaction is given by
\begin{align}
	\tilde{C}(i) = w Y(i),\label{consisol}
\end{align}
i.e., the canonical result that consumption is a constant fraction of income. Substituting $\tilde{C}(i)$ from eq.~\eqref{consisol} into eq.~\eqref{gen_utility} yields
\begin{align}
	U(\gamma; Y(i), C(j\mid i), c,w) &= (\gamma_i Y(i) - c\cdot(C(j\mid i) -  w Y(i) ))^w ((1-\gamma)Y)^{1-w}\label{gen_utility}
\end{align}

By the FOC we derive the optimal $\gamma_i$ as
\begin{align}
	\gamma_i &= \frac{wY(i) + (1-w) c (C(j\mid i) - wY_i)}{Y(i)},
\end{align} 
and consequently consumption by noting that $C(i) = \gamma_i Y(i)$ as
\begin{align}
	C(i) &= w Y(i) +  (1-w) \cdot c (C(j\mid i) - w Y(i)).\label{deriv_util}
\end{align}

Since we assume $w$ and $c$ to be constant for all agents, the consumption rule is entirely equivalent for all agents. This implies that the consumption of any agent $i$ derived here from a utility maximising framework is a  weighted sum of idiosyncratic and socially determined consumption with weights $w$ and $(1-w)$.\footnote{Note that this formulation is reminiscent of the consumption rule employed in the foundational paper by \cite{Frank2014} who just postulate and not derive the consumption function from utility maximisation, though.}

We want to emphasise that the derivation by utility maximisation is not the only conceivable way to arrive at such a consumption rule. Equivalently, the rule can also be interpreted as resulting from behavioural heuristics, where $w$ is the marginal propensity to consume out of current income and $c$ is the propensity to consume out of the consumption differential to the individual indexed $j$ without any reference to utility functions. In the terminology of evolutionary theories of consumption, the first term in~\eqref{deriv_util} would capture the `needs' of consumers with the latter term corresponding to the socially constructed `wants' \citep{witt2001, rengs2019}. In Post-Keynesian jargon, these terms would instead be interpretable as capturing the intuition of the absolute and relative income hypothesis \citep{palley2010}, respectively. The formal representation is in this sense neutral with respect to the subtleties of marginalist, evolutionary or Post-Keynesian interpretations and can thus be incorporated in those different frameworks without much difficulty.\footnote{In fact, even though social interdependencies seem now foreign to orthodox marginalist approaches, they were subject to large debates in the early days of marginalist consumer theory at the turn of the twentieth century \citep{bianchi2015}. In this sense, a heuristics-based approach might not be that far from a marginalist approach based on purposeful optimisation, with the two approaches not differing too much in the predicted consumption behaviour but only in the explanation of this behaviour.} 

Overall, intertemporal motives only feature within the model insofar as agents derive utility directly from saving in anticipation of future consumption and hence leave no role regarding expectation formation regarding potential future income gains or losses. However, our approach is also consistent with the permanent income hypothesis's central intuition: One can also understand the interpersonal consumption comparison presented here as a proxy for intertemporal ones. Individuals might look at others they perceive to lead overall parallel lives as themselves and take the observed consumption as a predictor for what is possible for themselves. In that way, our model explains why life-cycle models replicate the four stylised facts but necessitate implausible behavioural assumptions to do so: Individuals act as if they smoothed over future incomes while actually observing some of these potential future incomes in others. Arguably, merely observing consumption decisions rather than forming explicit (and model-consistent) expectations over a stochastic income innovation process appears to be much more plausible.\footnote{We thank Caverzasi Eugenio for pointing us to this alternative interpretation.}

Given the model mechanism, the income levels $Y(j\mid i)$ on which the consumption of $i$ (directly or indirectly) depends can be indexed by an ordering $\mathcal{R}_i = 0, 1, ... , d_i$ where $d_i$ is the distance of the most distant income to $i$ on which $C(i)$ depends. It follows that $Y (0\mid i) = Y(i)$, as the distance is then $0$. By recursive substitution, we can rewrite $C(i)$ as
\begin{align}
	C(i) &= \sum_{j=0}^{d_i} \large((1-w)\cdot c\large)^{j} \cdot (1- (1-w) \cdot c)\cdot~w\cdot Y(j\mid i)\\
	&= \sum_{j=0}^{d_i} \beta_{j} Y(j\mid i). \label{discounted}
\end{align}
The consumption of any individual $i$ is thus directly dependent on their own income but depends also on all other incomes to which they are (directly or indirectly) connected. Since $((1-w)\cdot c\large)$ is, per assumption, strictly below unity, the weight of a given distant income level $\large((1-w)\cdot c\large)^{j}$ decays (exponentially) in the graph distance $j = 0, 1, ..., d_i$ to the relevant agent. While more distant incomes are thus generally higher due to social comparisons being upward-looking, they are also given less weight. This implies that $C(i)$ is a linear combination of an exponentially distributed random variable $Y$ with varying weights and number of terms and thus, a so-called hypoexponential mixture \citep{li2019,yanev2020}. As we discuss in more detail in section~4, this directly implies two of the four stylised facts, namely, the approximate log-normality of expenditure distributions and the fact that they are robustly more homogeneous than income distributions.

Our homophilic graph formation process in conjunction with the upward-looking consumption rule thus formalises a central hypothesis by Thorstein Veblen explicitly at the micro level who conjectured that ''[...] our standard of decency in in expenditures, [...] is set by the usage of those next above us in reputability; until, in this way, especially in any community where class distinctions are somewhat vague, all canons of reputability and decency, and all standards of consumption, are traced back by insensible gradations to the usages and habits of thoughts of the highest social and pecuniary class`` \citep[p. 72]{veblen2001}. While the homophilic graph formation guarantees that agents only react to consumption of those that are close to them regarding their social status, eq.~\eqref{discounted} shows that the consumption of the poor might indirectly depend even on the consumption of the richest agents through the network.

\section{Results}\label{sec:Results}
The results we present are twofold: Firstly, we show that our parsimonious model setup embedded within realistic social networks can replicate the known stylised facts of empirical expenditure distributions at the micro-level. On the one hand, these results testify to the validity of our model assumptions. On the other hand, they also serve as proof of concept to show that a static setting may replicate empirical expenditure patterns without resorting to unobservable stochasticity, strong assumptions about credit markets and individual rationality or assuming any heterogeneity in individual consumption functions ex-ante. Secondly, we spell out the implications for savings and inequality at the macro level. In particular, we show that our model can generate a large variety of different elasticities of savings with respect to inequality, in line with the highly ambiguous empirical literature. 

\subsection{Micro Level Patterns}
\begin{figure}[!t]
	\begin{center}
		\includegraphics[width=.8\linewidth]{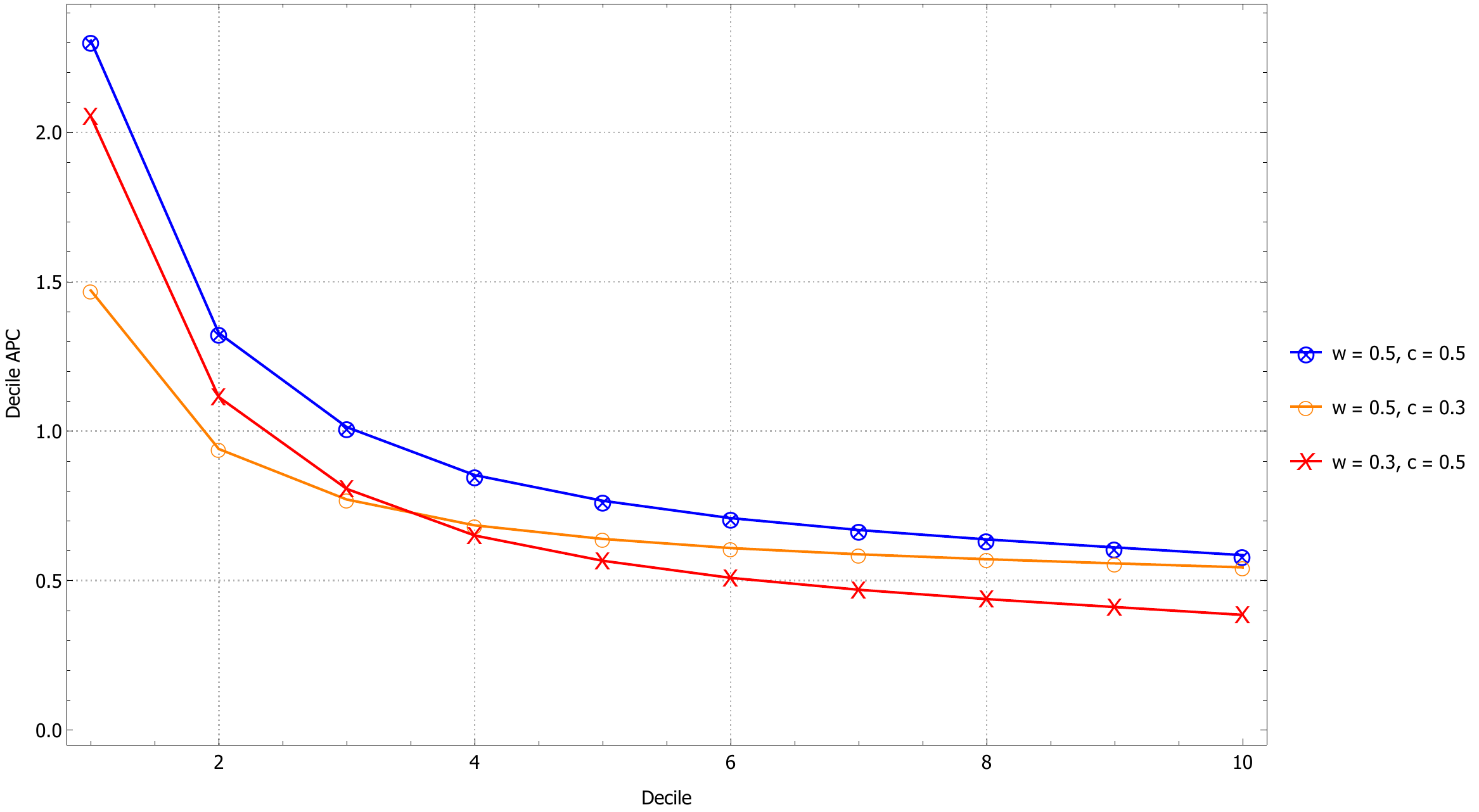}
		\caption{The figure plots the APCs per income decile for different parametrisations and with homophily strength $\rho = 4$ for a single simulation run each. Lines are visual aids only.}\label{fig:decapc}
	\end{center}
\end{figure}

As laid out in section~2, we consider four stylised facts for validation with varying levels of granularity. Stylised facts (i) to (iii) describe qualitative features of expenditures that are, in principle, consistent with many different functional forms describing the distribution of expenditures. The fourth stylised fact is more stringent and imposes log-normality as the functional form. Even though log-normality is a more demanding requirement, it does not itself imply stylised facts (i) to (iii), which we, therefore, consider separately for full validation \citep{fagiolo2019}. The first stylised fact concerns the decrease of individual APCs in income levels. Indeed, our model replicates this finding for all levels of social consumption without imposing different consumption propensities ad hoc. This follows directly from social consumption being purely upward-looking. If one is richer than all their link-neighbours, they do not need to catch up to anyone's higher consumption, and hence their consumption is entirely idiosyncratic. However, social consumptfion accumulates as it is passed down within the network from richer to poorer individuals. Thus, poorer individuals tend to exhibit higher social consumption above their idiosyncratic consumption, increasing their personal APC. Figure~\ref{fig:decapc} shows the APC schedules per decile for different parametrisations and demonstrates that decile APCs are indeed declining, in line with the empirical evidence. Note that this result is endogenous to our model and not imposed \emph{ex ante}, as is frequently postulated in class-based macroeconomic models \citep[cf.][for are recent example]{rengs2019}.

As Figure~\ref{fig:decapc} also shows, though, the precise form of this decay depends on the specific parametrisation. The $w$ parameter determines the weight of idiosyncratic consumption. Since idiosyncratic consumption is relatively more important for the richest, $w$ is also most relevant for the level of the total consumption of the richest decile, as is apparent from the fact that the blue and orange APC schedules with equal $w$ but different $c$ in Figure~\ref{fig:decapc} almost coincide for the richest decile, whereas the red and orange schedules with equal $c$ but different $w$ do not. In contrast to that, the $c$ parameter captures the strength of the `catching-up' behaviour, which manifests itself in the steepness of the curve, as is also readily visible from comparing the orange and red APC schedules.\footnote{We chose these particular parametrisations, since, for  $\rho = 4$, they appear to capture the behavior of empirical APCs reported in \cite{clementi2017} reasonably well. Decreasing $\rho$ tends to increase the reported effects since the income difference to the richest observed individual within a perception set tends to increase with decreasing $\rho$ by assumption, leading to stronger `catching-up'. Simulation results verifying this finding are available upon request.} Regardless of the parametrisation leading to different functional forms of the APC schedules, they are, however, all declining in income decile, replicating stylised fact (i).

The most prominent fact regarding empirical expenditure distributions was the finding by \cite{kuznets1942} that aggregate APCs stay approximately constant for changes in aggregate income. Indeed, this was one of the major factors contributing to the paradigm change from the absolute to the permanent income hypothesis, as the absolute income hypothesis in its affine-linear variant exhibits decaying aggregate APCs in total income \citep{palley2010}. The invariance of aggregate APCs $C/Y$ from changes in $Y$ follows trivially from the recursive solution to $C(i)$ in eq.~\eqref{discounted}. Assume a proportional change to $Y$ by $\omega \in \mathbb{R}^+$, such that $\tilde{Y}(i) = \omega Y(i)~~\forall i$ and  $\tilde{Y} = \omega Y$. Thus, $\omega$ is independent of $Y(i)$ and does not change orderings $\mathcal{R}_i$. Consequently, since $\omega$ cancels out, total consumption is given by the following:
\begin{align}
	\tilde{C}  &=  \sum_{i=1}^n \sum_{j=0}^{d_i} \beta_{j}  \omega Y(j\mid i) = \omega \sum_{i=1}^n \sum_{j=0}^{d_i} \beta_{j} Y(j\mid i)~\text{and therefore}\\
	\frac{\tilde{C}}{\tilde{Y}} &= \frac{\omega \sum_{i=1}^n \sum_{j=0}^{d_i} \beta_{j} Y(j\mid i)}{\omega \sum_{i=1}^n   Y(i)} = \frac{C}{Y}.
\end{align}
As a result, aggregate APCs are invariant to any positive proportional change to income, i.e., distribution-preserving changes. Whenever inequality by one of the usual measures increases, aggregate APCs increase, too, as we show in the subsection below.

As a third stylised fact, we considered that the distribution of consumption expenditures is more homogeneous than the income distribution. In our specification, this is necessarily also true for $w < 1$,\footnote{For $w = 1$, we would simply recover a (rescaled) version of the exponential income distribution for expenditures, and the recursive solution in eq.~\eqref{discounted} would be undefined.} since by the recursive solution in eq.~\eqref{discounted}, expenditure levels are a weighted sum of income levels, themselves following an exponential distribution, with varying weights and number of terms. This characteristic implies that the distribution of $C$ is \emph{hypoexponential}, with a coefficient of variation strictly smaller than unity, while the exponentially distributed income levels exhibit a coefficient of variation (asymptotically) equal to unity \citep{li2019,yanev2020}. Intuitively, since social consumption accumulates in a cascade down the income distribution, richer individuals are relatively unaffected by status concerns, while the cumulative effect on poor individuals is much higher. Thus, increasing the level of status consumption by decreasing $w$ or increasing $c$ tends to equalise the expenditure distribution since the relatively higher status consumption of the poorer lets their consumption approach the expenditures of the rich that are not so much affected by status concerns.

\begin{figure}[!h]
	\begin{center}
		\minipage{0.4\textwidth}
		\includegraphics[width=\linewidth]{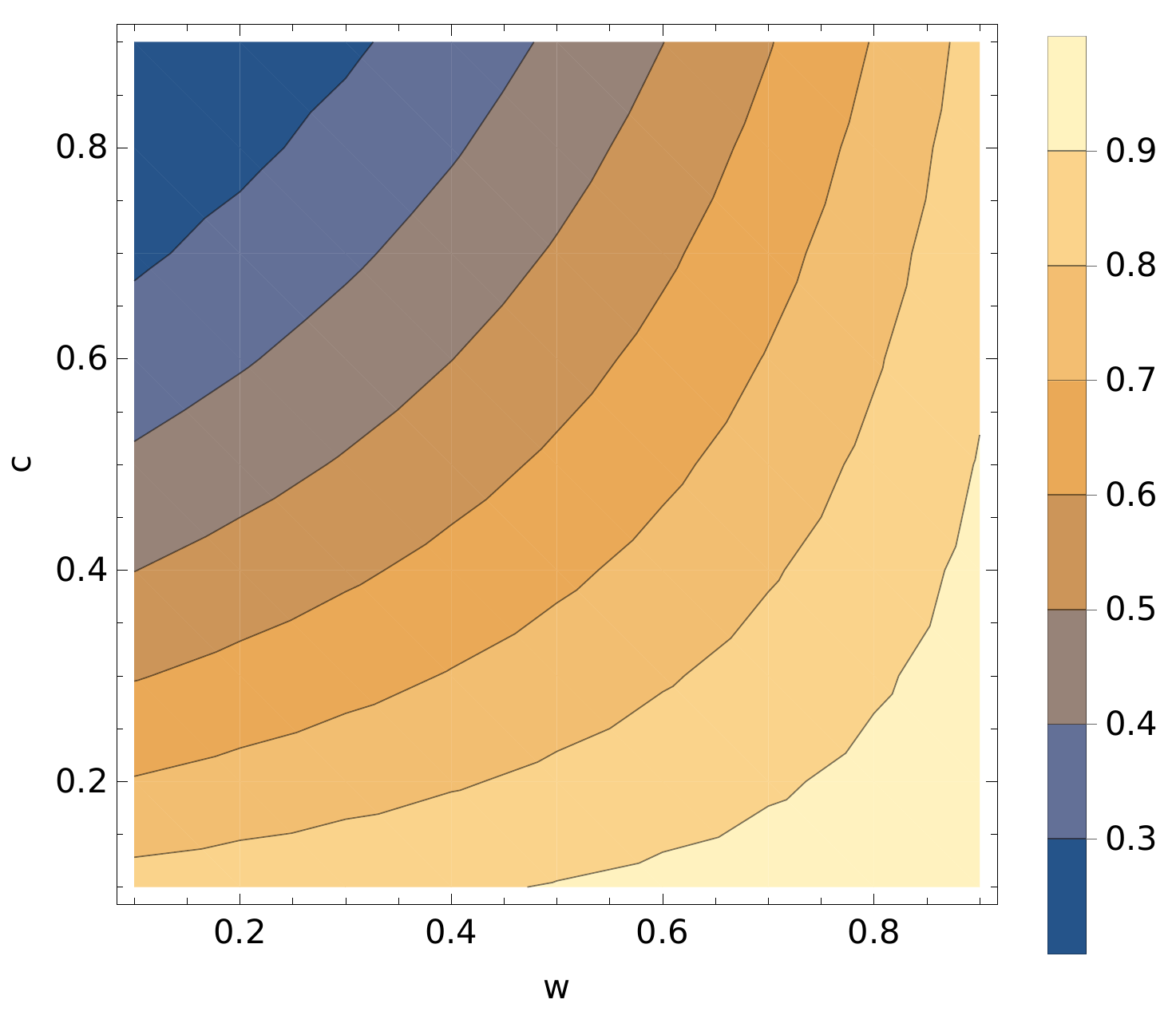}
		\caption{Ratio of the coefficients of variation for the expenditure and income distribution for the whole parameter space of $w$ and $c$ and with $\rho = 0.5$, averaged over $100$ simulation runs.}\label{fig:covrho05}
		\endminipage        \hspace{0.5cm}
		\minipage{0.4\textwidth}
		\includegraphics[width=\linewidth]{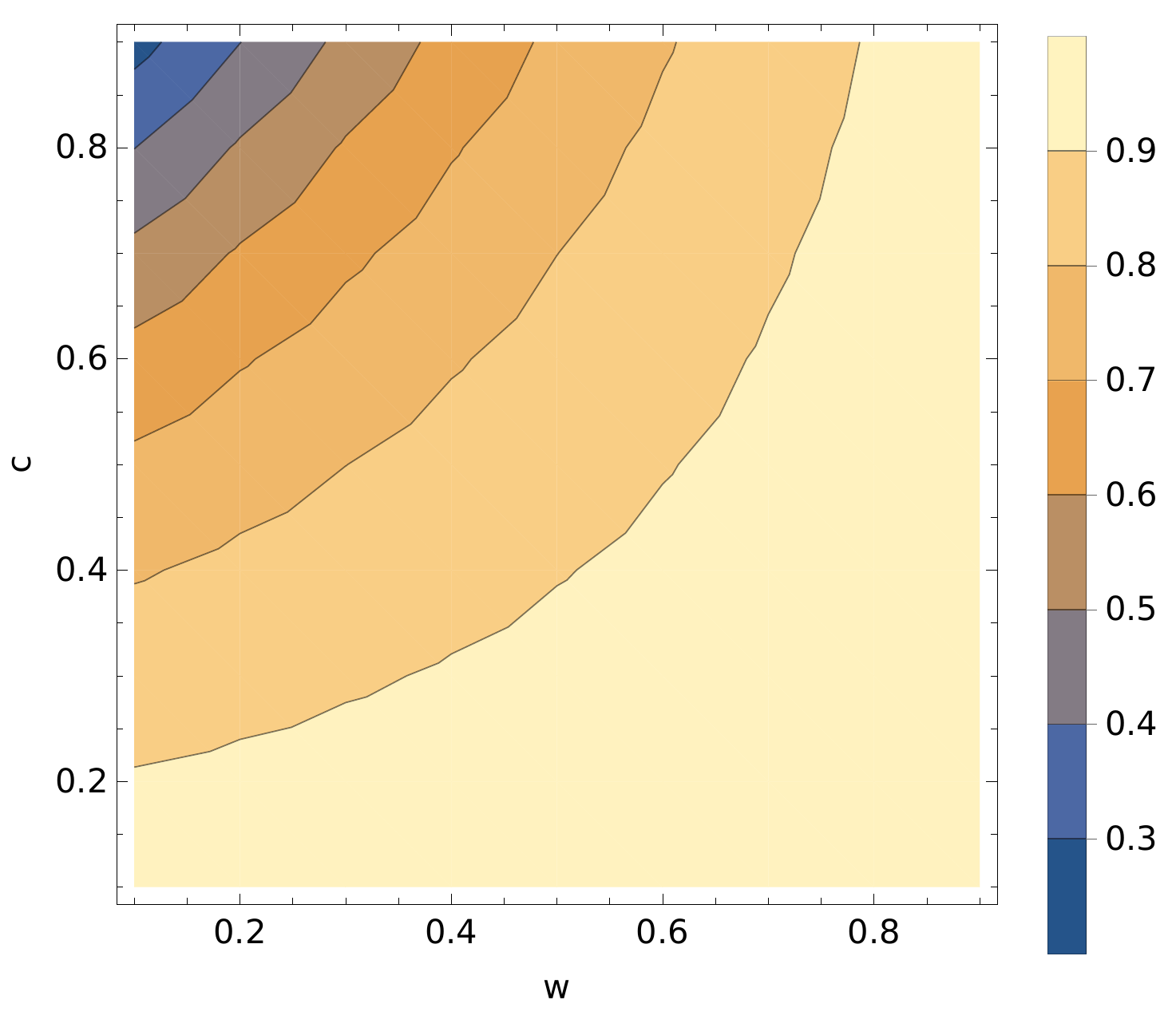}
		\caption{Ratio of the coefficients of variation for the expenditure and income distribution for the whole parameter space of $w$ and $c$ and with $\rho = 4$, averaged over $100$ simulation runs.}\label{fig:covrho4}
		\endminipage
		
	\end{center}
\end{figure}

This is also what we find for our numerical simulations, here for $100$ runs each, as we show in Figures~\ref{fig:covrho05} and~\ref{fig:covrho4} for the parameter combinations of $w$ and $c$. For all cases, the coefficient of variation is strictly below the one for the income distribution, replicating stylised fact (iii). However, the equalising tendency is much less pronounced for high levels of homophily, as can be seen in Figure~\ref{fig:covrho4} because the much more homophilic link formation tends to decrease the maximum income level each individual observes. Hence, homophily tends to hinder the transmission of social consumption.

\begin{figure}[!h]
	\minipage{0.45\textwidth}
	\includegraphics[width=\linewidth]{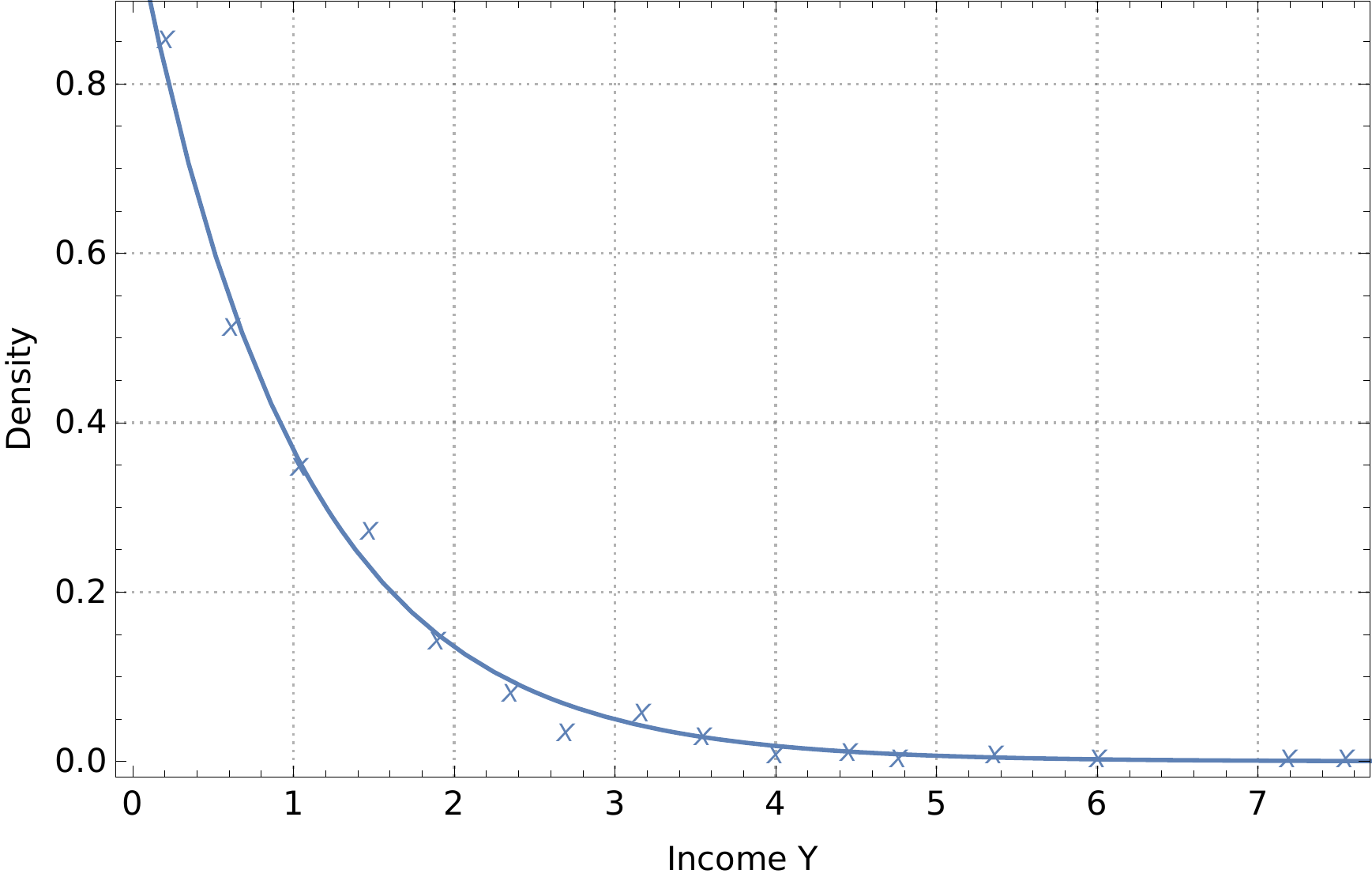}
	\caption{\footnotesize Empirical Density of Personal Incomes $Y(i)$.}\label{fig:inc}
	\endminipage\hfill
	\minipage{0.45\textwidth}
	\includegraphics[width=\linewidth]{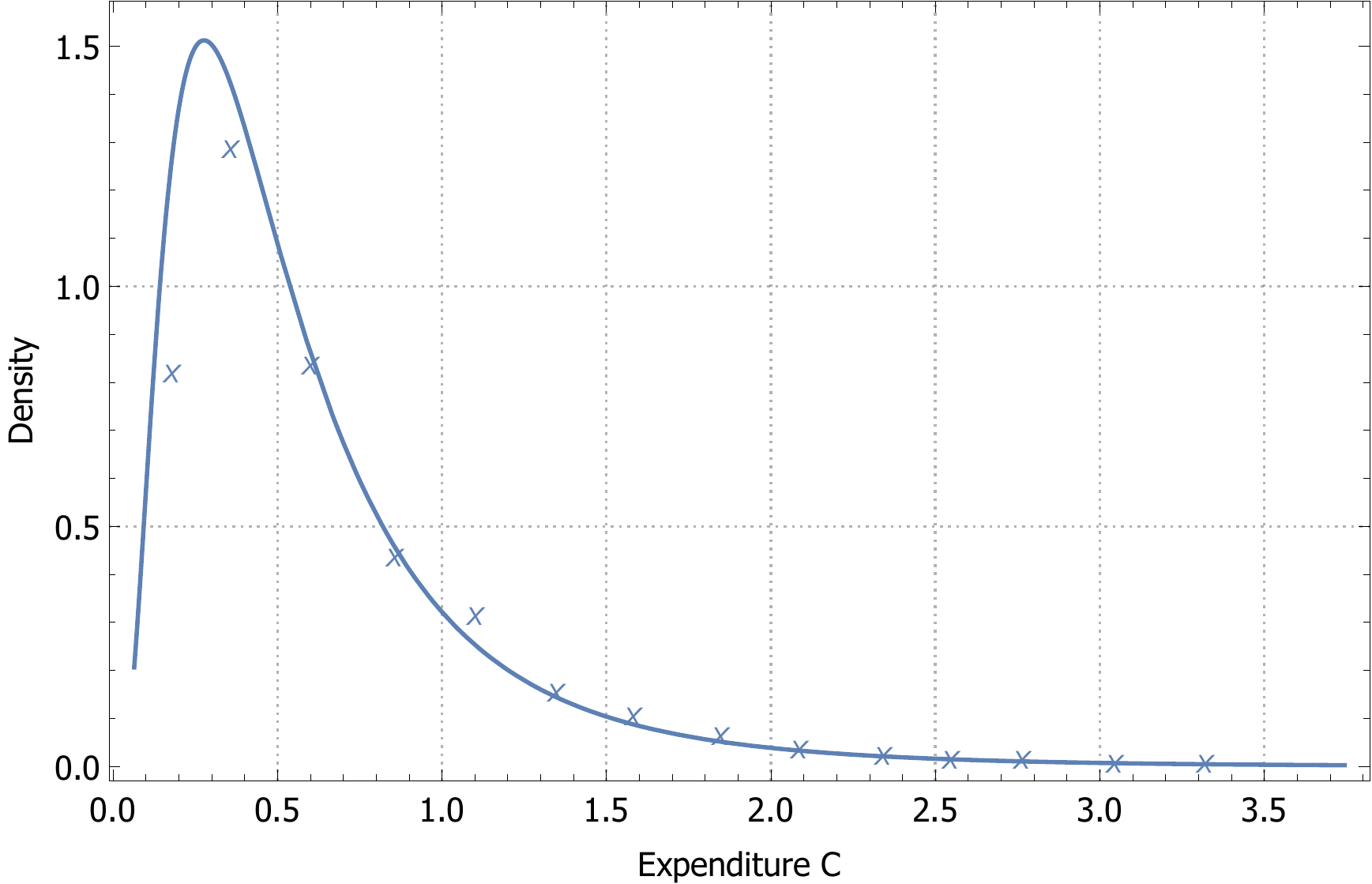}
	\caption{\footnotesize Empirical Density of Personal Expenditures $C(i)$ for a single simulation run with $w = 0.5$, $c = 0.3$ and $\rho = 1$.}\label{fig:exp}
	\endminipage\hfill
\end{figure}

Finally, we also find that our specification can map exponential income distributions to log-normal expenditure distributions, even though both come from completely different distributional families. The reason behind this finding is simply that the hypoexponential mixture characterising the expenditures is well approximated by a log-normal if it exhibits sufficient skewness, but the underlying rate parameters $\lambda$ of the exponential are not too heterogeneous. A representative example is shown in Figures~\ref{fig:inc} and~\ref{fig:exp}.

Per assumption, the pre-validated income distribution follows an exponential distribution with rate parameter $\lambda = 1$. However, the expenditure distribution is extremely well fit by a log-normal distribution, replicating stylised fact (iv). Note that this represents an entirely new generating mechanism for log-normal distributions and does not rely on any unobservable income innovation process or model-consistent expectations as are typically used in mainstream macro models. This is a relevant finding in and of itself since the log-normality of consumption expenditures is typically taken as evidence that consumption cannot be a function of \emph{current} income and thus needs to depend on \emph{permanent} income subject to a stochastically multiplicative growth process generating the log-normal functional form \citep{battistin2009}. Our model exercise demonstrates that this is not necessarily true. A parsimonious model based on current income and simple behavioural rules of thumb can simultaneously replicate this stylised fact without facing the theoretical and empirical shortcomings of models in the rational expectations tradition.

\begin{figure}[h]
	\minipage{0.45\textwidth}
	\includegraphics[width=\linewidth]{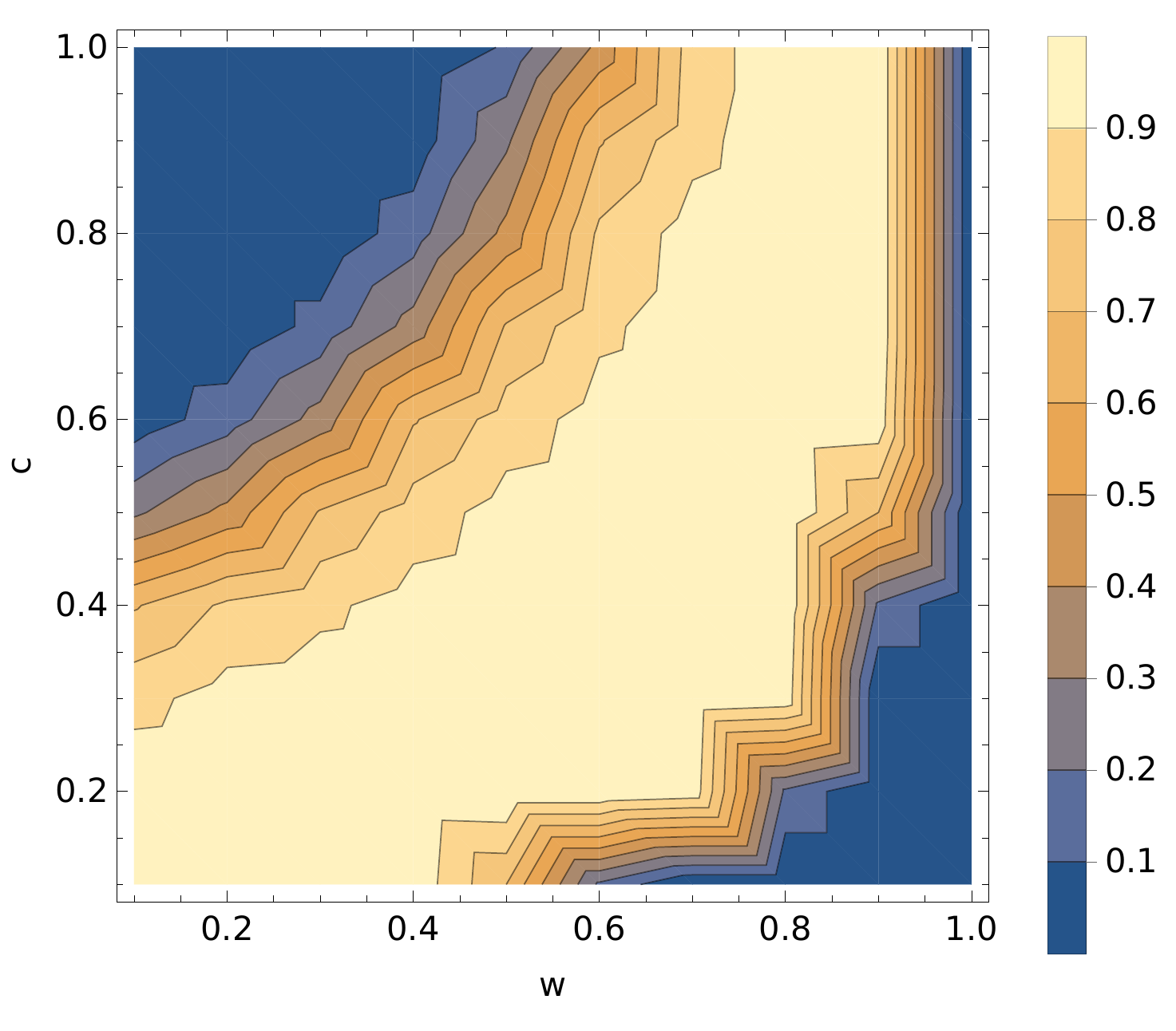}
	\caption{\footnotesize Fraction of expenditure distributions for which a standard Kolmogorov-Smirnov test cannot reject the assumption of log-normality at a 5\% significance level for $\rho = 0.5$ with $100$ iterations.}\label{fig:lnrho05}
	\endminipage\hfill
	\minipage{0.45\textwidth}
	\includegraphics[width=\linewidth]{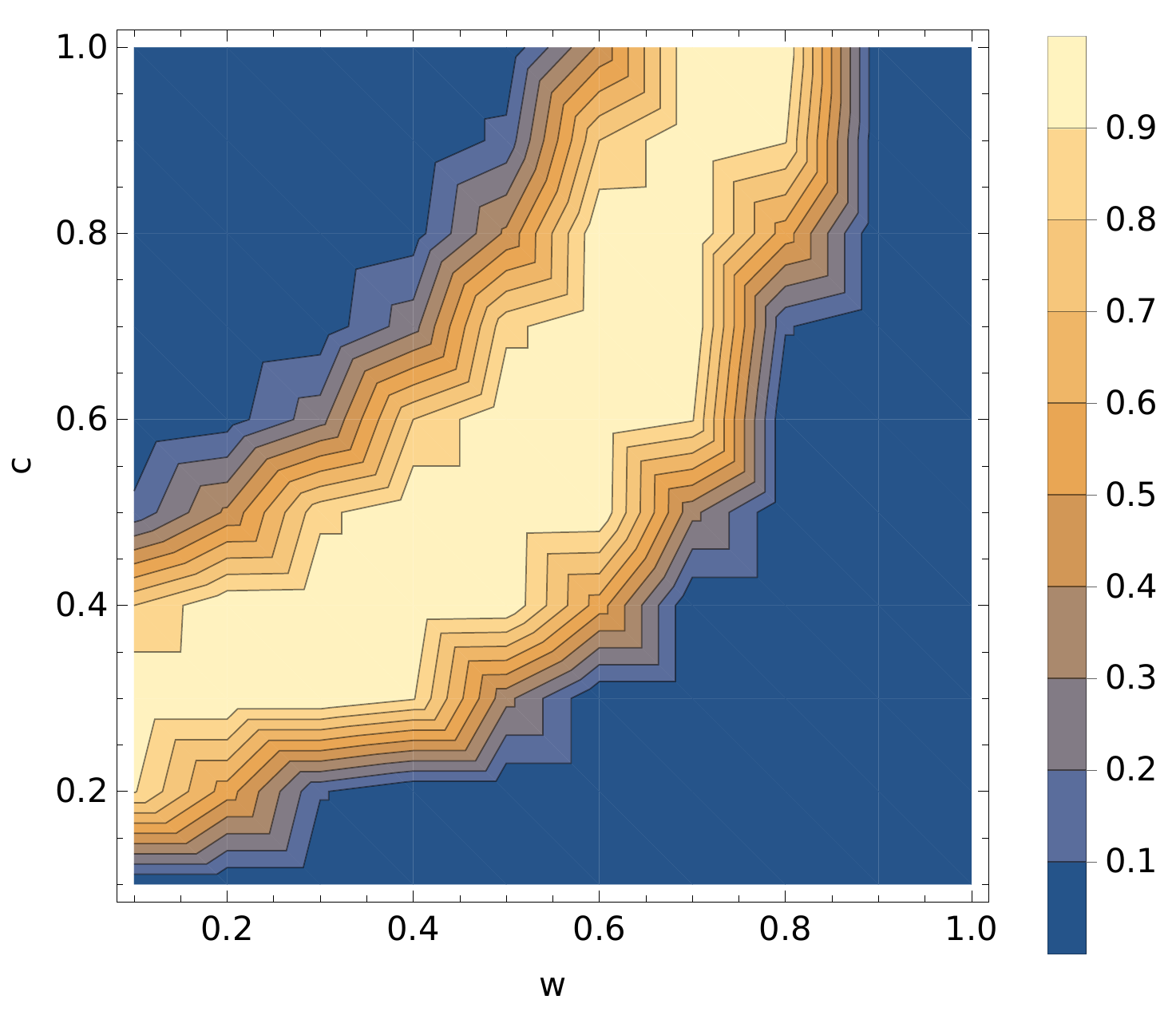}
	\caption{\footnotesize Fraction of expenditure distributions for which a standard Kolmogorov-Smirnov test cannot reject the assumption of log-normality at a 5\% significance level for $\rho = 4$ with $100$ iterations.}\label{fig:lnrho4}
	\endminipage\hfill
\end{figure}

Log-normality holds for large parts of the parameter space, as we show in Figures~\ref{fig:lnrho05} and~\ref{fig:lnrho4}. The approximation only breaks down for very high levels of idiosyncratic consumption and $w$ approaching unity, where we recover a rescaled version of the initial exponential distribution and very high levels of social consumption with low $w$ and high $c$, where the expenditure distribution becomes too symmetric to account for the skewness of the log-normal. For high levels of homophily $\rho$, the parameter space for log-normality becomes much narrower due to the emergent segregation that weakens the transmission of status consumption. In light of these findings, it seems entirely unsurprising that many of the empirical studies we reviewed in section~2 find significant deviations from log-normality in expenditure distributions, which one can understand as arising from variation in consumption baskets' salience of expenditures as well as differences in social segregation. Both the heterogeneity in the salience of consumption goods \citep[for a large scale study]{solnick2005, heffetz2011} and the variation in (e.g., spatial) segregation \citep{toth2021} are well documented in the literature, in agreement with the varying results on the log-normality of expenditures across studies. Perhaps more surprisingly, we find that our stylised consumption function embedded within realistic social networks can replicate the relevant empirical phenomena in the literature on empirical expenditure patterns, testifying our modelling approach's  validity. 

\subsection{Inequality and Aggregate Savings}\label{subsec:aggsav}
So far, the model assumed exponentially distributed incomes with a constant theoretical Gini coefficient of $G = 0.5$, irrespective of its particular parametrisation. To examine the impact of \emph{changing} inequality on consumption and savings, we therefore now assume log-normality of individual incomes to generate variation in inequality through changing the dispersion parameter $\sigma$ of the log-normal income distribution by which the model is initialised. We examine different degrees of homophily and let social networks adjust endogenously to input inequality, thereby highlighting two different findings: Firstly, there exist strong non-linearities in the relation of inequality and savings whenever perception networks mediate actual inequality. Secondly, the time scale of adjustment might be important. We only provide comparative statics here and report results after all adjustments regarding perceptions have taken place but hint at possible mechanisms in the time domain as well.

\begin{figure}[!t]
	\begin{center}
		\includegraphics[width=.8\linewidth]{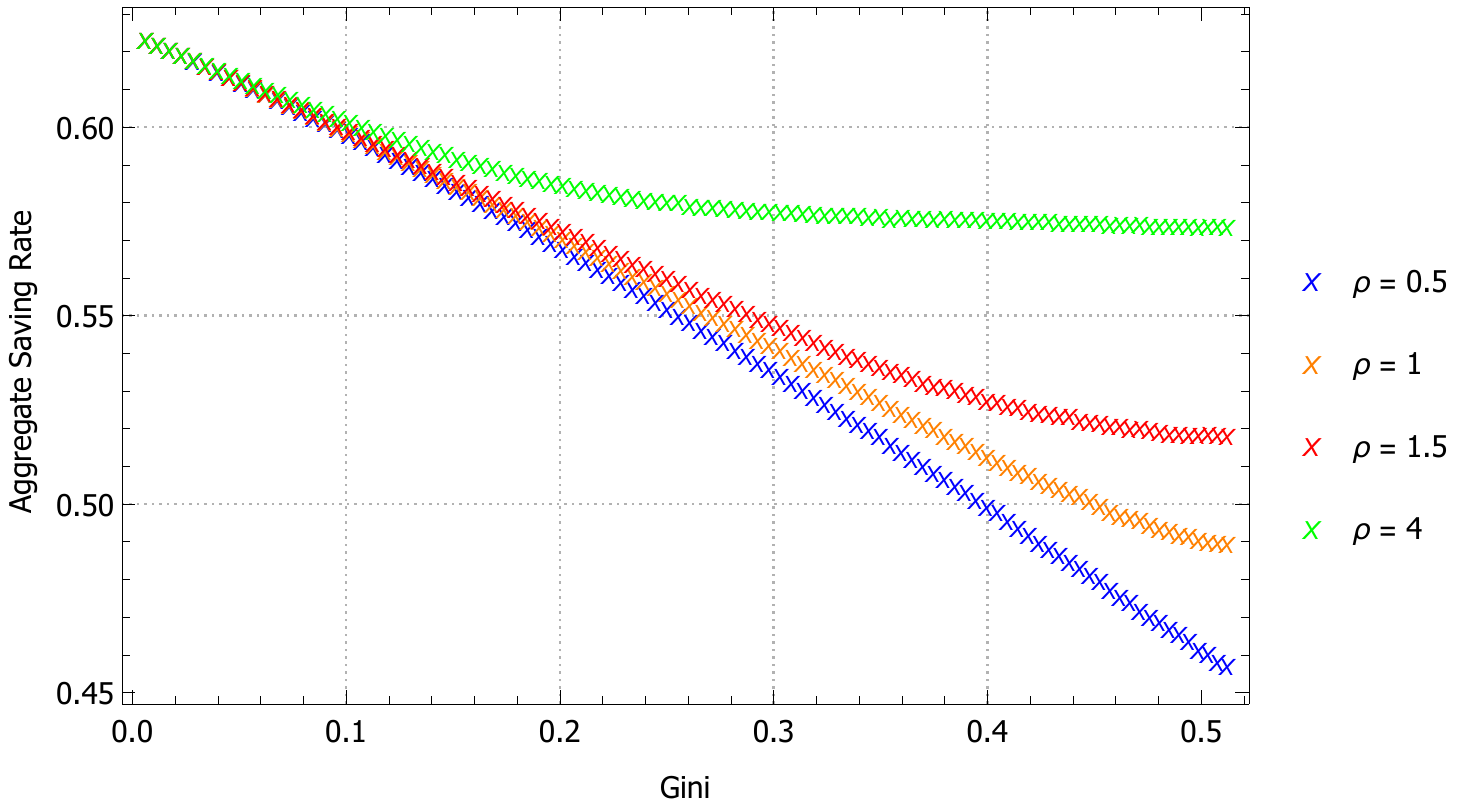}
		\caption{Aggregate saving rate (aggregate savings over aggregate income) as a function of inequality, measured by the Gini coefficient with $w = c = 0.5$ and $b = 0.75$, averaged over $100$ simulation runs.}\label{fig:aggaps}
	\end{center}
\end{figure}

Figure~\ref{fig:aggaps} depicts aggregate saving rates as a function of income inequality, measured by the Gini coefficient, and for varying levels of income homophily. As expected within a model of purely upward-looking consumption, savings unanimously decrease in inequality, as lower-income classes try to emulate the higher expenditures of the rich caused by their disproportionate gains in income. We are thus able to replicate the main finding of the canonical model of `expenditure cascades' by \cite{Frank2014} within plausible and endogenously evolving social networks. However, we also find that effect sizes are both state-dependent for the very same behavioural parameters and decrease with homophily because for the given graph-generating process, two counteracting effects mediate the effect of actual inequality on perceived inequality \citep{schulz2022}. Firstly, holding the network-topology constant, increases in aggregate inequality also increase inequality \emph{within} perception groups and thus decrease savings. Secondly, the network topology endogenously mitigates an increase in inequality, too: This increasing inequality leads to higher segregation, which means higher inequality \emph{between} groups but decreasing inequality within every given group. A priori, it is unclear what the relative strengths of both effects are, which we explore by simulation.

For (implausibly) low degrees of inequality, the four trajectories for $\rho = 0.5, 1, 1.5$ and $4$ almost appear to coincide, as the between-effect is negligible here, while the within-effect is fully passed through to savings. From a Gini coefficient $G = 0.1$ onwards and now within the empirically relevant range \citep{bofinger2019}, the trajectories begin to deviate, with still strongly negative effects for the low-homophily regimes but vanishing savings elasticities within the highly homophilic regime. This is apparent from the decaying schedules for $\rho = 0.5, 1$ and $1.5$ in contrast to the plateau for $\rho = 4$. Methodologically, this finding thus points to a paradox of aggregation: With completely identical behavioural rules at the micro-level, one can generate vastly different aggregate savings elasticities at the macro level. Reminiscent to \cites{kirman1992} foundational demonstration, the properties even of identical micro agents do not translate to equivalent properties of a `representative agent' whatsoever, as the nonlinear macro behaviour in Figure~\ref{fig:aggaps} in contrast to the identically linear micro consumption rule shows. Albeit very different in detail, this result testifies to the relevance of the complex interaction of meso-level rules for aggregate consumption that \cite{foster2021} finds to be of crucial relevance for aggregate consumption patterns.

This methodological finding is not only a theoretical curiosum but also bears material implications: Most importantly, it might in part explain why the micro literature appears to confirm the relevance of the conspicuous consumption channel, while the macro literature remains at best ambiguous on this issue. It might very well be the case that all individuals engage in conspicuous consumption without this being observable using only macro-level aggregates, whenever there is an endogenous selection of reference groups. Indeed, the scarce extant empirical literature suggests both that inequality increases (geographical) segregation \citep{toth2021} and that (geographical) segregation decreases conspicuous consumption \citep{bertrand2016}. Policies aimed at decreasing segregation might thus carry unintended consequences in simultaneously increasing conspicuous and potentially wasteful consumption. 

Finally, our results are also a cautionary tale against the pitfalls of pooling macro data for analysis without acknowledging the potential time- and cross-sectional variation of inequality effects on segregation and perceptions. Indeed, it appears likely that the time-scale of the homophily mechanism outlined in section~3 crucially depends on the mobility of actors that enables homophilic choice or induced homophily in the first place \citep[for a recent empirical study]{thomas2019}. With low mobility, the transitory adjustment to the new stationary state we depict in our comparative statics exercise in Figure~\ref{fig:aggaps} might take sufficiently long to be of macroeconomic relevance. In contrast, adjustment might be very fast for high degrees of mobility. The significant short-term elasticities, e.g. in \cite{petach2021} can therefore be reconciled with the insignificant long-term effects \cite{wildauer2018} find by considering heterogeneous mobility in contrast to ad-hoc assumptions regarding behavioural parameters.

Furthermore, the degree of homophily itself appears to be varying across and within countries \citep{Mcpherson2001,cepic2020}. Consequently, pooled estimates might obscure unobserved cross-country heterogeneity both in adjustment speed and strength, which might help to explain the non-monotonicity \cite{bofinger2019} find. Therefore, our model also calls for better data to capture these effects, either by controlling for, e.g., residential or occupational segregation or by directly including measures of perceived incomes. Unfortunately, these attempts are most of the time hindered by data that is much too coarse-grained for proper estimation, with the geographical location only being available at the state or county level \citep{bertrand2016} and perceived incomes being reconstructed from data on a rather crude ordinal scale \citep{Choi2019} or based merely on ordinary workplaces \citep{degiorgi2020}. Merging the relationship network, including friends and family, besides coworkers with the theoretically relevant socioeconomic data is rarely achieved in practice but would be an important step to further bridge social network and consumption theory \citep{depaula2017}.

\section{Discussion}\label{sec:Discussion}
Our goal was to provide a tractable evolutionary alternative to orthodox theories of consumption based on Euler equations and intertemporal optimisation. In particular, we built on the distinction between `needs' and `wants' \citep{witt2001} and integrated social norms through a perception network \citep{witt2010}. Our results suggest that this simple consumption rule based merely on current income and explicit observables can replicate the same set of major stylised facts as the orthodox formulation based on intertemporal consumption smoothing. In this sense, the two consumption theories appear to be observationally equivalent, and thus, even microdata on expenditure distributions might be unable to discriminate between the two. Theory, even in the form of carefully calibrated macro models, might be \emph{underdetermined} in this regard \citep{quine1975}. Intertemporal motives might matter for consumption, but the log-normality of expenditures is perhaps not an as decisive argument for consumption smoothing as it is often perceived to be \citep{battistin2009}. Methodologically, intersubjective rather than intertemporal consumption rules might also be appealing since they build on much less restrictive assumptions. Intertemporal optimisation essentially implies both quantifiable income risk rather than fundamental uncertainty in the form of a well-defined probability distribution over income shocks and demands that this distribution is learnable or transparent for consumers \citep{menz2010}. By contrast, social consumption rests on cross-sectional observables made explicit by the perception network and therefore requires agents to be mere observers rather than econometricians which strikes us as a more plausible portrayal of human behaviour \citep{nelson2010}. At the same time, our micro-level mechanisms are consistent with the permanent income hypothesis, too. By acting on what they see in others they perceive to be similar to themselves, individuals behave as if they were acting on expectations about their own lives. Hence, using social networks relying on interpersonal consumption comparisons might be attractive for future research as a proxy for intertemporal considerations and its own sake because conspicuous consumption itself is an empirically well-grounded phenomenon.

Our second major result addresses the aggregation of individual consumption within the perception network. We find that, paradoxically, aggregate consumption might react far less elastic to changes in income than each individual consumption rule considered separately would indicate, even when all individuals unanimously follow exactly the same rule. This apparent contradiction follows from the endogenous selection of consumption reference groups. Apart from increasing aggregate consumption, inequality also increases segregation and therefore mitigates its initial consumption-enhancing effect. Naturally, this new channel sheds light on recent puzzles concerning the apparent divide between the micro and macro strands of the empirical literature on this matter. In particular, we explicate the potential micro origins of the institutional channel that \cite{behringer2018b} and \cite{ascione2021} find to be of crucial relevance to explain the empirical cross-country heterogeneity and time variation: In addition to the ease of access to credit or consumerist social norms, also the social network of everyday interactions might be an important determinant for relative importance of expenditure cascades. The growing literature on `growth models' within comparative political economy might thus also benefit from considering institutional differences in social segregation \citep{behringer2019}. 

Moreover, our results point to a relevant trade-off for policy-makers: According to our model results, policies aimed at reducing occupational or residential segregation might also increase conspicuous consumption that might be environmentally wasteful \citep{howarth1996} and contributes to the destabilising build-up of private debt \citep{vantreeck2014}. To offset these effects, our framework suggests complementing such policies of integration by appropriate taxation or behavioural nudges that are disincentivising status consumption \citep{antinyan2020}, or, in the language of our model, decrease $c$. Note that this channel should be independent of our specific choice of consumption rule and qualitatively carries over to models with different utility functions, in particular, ratio instead of additive utility \footnote{Cf. the discussion in \cite{alpizar2005}.}, as long as there exists an upward-looking social consumption component as well as perception networks that are homophilic and adjust endogenously.

The choice in favour of a unanimous consumption rule based on comparisons is a deliberate one for two reasons: Firstly, our parsimonious model's internal validation was unproblematic as we could monitor the whole parameter space of all exogenous variables and did not detect any inexplicable irregularities. Secondly, we keep individual behaviour intentionally simple to show that the \emph{interaction} of agents is sufficient to generate empirical heterogeneity in micro observables and a large variety of macro-level effects. By applying a minimal model of consumption, we identify a (possible) minimal set of necessary assumptions \citep{grune2009}. Hence, our model constitutes a how-possibly explanation of the empirically observed consumption patterns. However, the explanans can be true in the real world and the cause for the observed empirical fact: Our proposed mechanism fulfils the minimum conditions for a good epistemically possible how-possibly explanation formulated by \cite{gruene2021}. That our simulation results are in ``qualitative agreement with empirical macrostructures'' \citep[p. 771]{fagiolo2019}, namely successful replication of the stylised empirical facts, affirms the external validity of the model. Put differently, we develop a specific parallel reality \citep{sugden2009} that features generating mechanisms for empirical findings in our reality and hence our results present a candidate explanation for the stylised empirical facts \citep{Epstein1999}. Consequently, there may be different, more adequate, parallel realities featuring either these or even better mechanisms, despite to the best of our knowledge, there are no existing models that fulfil these characteristics. Furthermore, our proposed framework allows for arbitrary changes in effect sizes and micro patterns even without social consumption. To achieve this, one simply needs to alter individual consumption parameters ex-ante, e.g., by letting poorer individuals idiosyncratically consume a larger portion of their income and, as a corollary, increase the effect of inequality on aggregate consumption. Finally, both the principle of homophilic choice and social consumption effects are well established in the extant empirical literature, further testifying to the resemblance \citep{maeki2009} between our model and reality. Therefore, our model reconciles technical verification with both input and descriptive output validation which \cite{grabner2018} considers a rare feature of models. We are confident that the how-possibly explanation given by our model is preferable to ad-hoc choices and much more general or could at least be straightforwardly extended in this direction.

Nonetheless, the minimal nature of our model points to several limitations and extensions. Most notably, in its current bare-bones form, the model cannot study any issues of liquidity constraints that are frequently found to be of crucial relevance, especially for low-income households \citep{dynan2004} or spill-overs of private consumption to other sectors that would, at a minimum, require the model incorporate general equilibrium effects and the model to be stock-flow consistent \citep{rengs2019}. Any supply-side factors or advertising activity to socially mediate `wants' following \cite{witt2010} are also absent within our model that might nevertheless be of importance to explain the \emph{changes} within consumption baskets and habits over time that \cite{foster2021} identifies. In particular the stock-flow consistent framework by \cite{rengs2019} or similar approaches could serve here as a starting point to incorporate our considerations within a full-fledged macro model.

Generally, links are modelled abstractly and incorporate various types of personal or professional connection layers as proximity of incomes. Future work could disentangle this conglomerate and represent specific ties more explicitly to focus on a particular connection layer, e.g., connections between co-workers implying network communities that represent job categories. Early contributions to development economics on analogous `demonstration effects' have emphasised the role of international emulation of consumption patterns \citep{nurkse1953}, while our perspective is much more granular and focuses on individuals. Exploring the interplay of aggregate frames of reference between countries and emulation of individual consumption within them might prove to be instructive to investigate cross-country convergence since the two could reinforce each other.

Furthermore, since the model is static, time does not play any role within the model per construction. \cites{simon1962} classic cautions us that the dynamic adjustment of different complex systems in response to exogenous disturbances might also operate on vastly different time scales, dependent on its decomposability properties. In this sense, the persistence of the effect of inequality on savings is a function of the time difference between consumption adjustment and network reshuffling. A possible remedy for this could be to explicitly model both consumption decisions, constrained by the availability of credit, and homophilic choice, constrained by mobility, within a full-fledged macro model. One promising avenue for empirical determination of the relevant time scales could be exploiting the time-variation in urbanisation rates. \cite{veblen2001} conjectured already in $1899$ that cities are a powerful tool for diversifying perceptions, since within them, ``the mobility of the population is greatest'' \citep[p. 66]{veblen2001}. The yet relatively scarce empirical evidence for Vietnam and Central and Eastern Europe suggests that perceived income diversity is indeed increasing with urbanisation \citep{mahajan2014, binelli2016}. Apart from theoretical models, we hope to invite empirical studies, especially at the intersection of urban or spatial economics, inequality research and consumer theory.

\backmatter

\bmhead{Supplementary information}
The NetLogo code for the model and various extensions as well as extensive documentation following the ODD protocol can be found at \href{https://github.com/mayerhoffer/Inequality-Perception}{https://github.com/mayerhoffer/Inequality-Perception}.

\bmhead{Acknowledgments}
Daniel Mayerhoffer's contribution was funded by the Deutsche Forschungsgemeinschaft (DFG, German Research Foundation) - 430621735. Furthermore, financial support for the paper by the University of Bamberg through the Fres(c)h grant no. 06999902 is gratefully acknowledged. We would also like to express our gratitude to Yunus Aksoy, Jan Behringer, Domenico Delli Gatti, Anna Gebhard, Gabriel Lozada, \"Ozlem Onaran, Benjamin Tippet, Rafael Wildauer and the participants of the research colloquium at the University of Utah as well as the 12th Annual PhD Student Conference of the Post-Keynesian Economics Society, especially to our designated discussant Caverzasi Eugenio, the participants of the 25th conference of the Foundation for Macroeconomics and Macroeconomic Policies and the participants of the Workshop on Economics with Heterogeneous Interacting Agents 2021 for many helpful comments and suggestions that greatly improved our paper and sharpened our focus. Finally, we also thank Marius Theiler and Carsten K\"allner for their able research assistance. All remaining errors are, of course, ours.

\section*{Declarations}
%

\begin{itemize}
\item Funding: Daniel Mayerhoffer's contribution was funded by the Deutsche Forschungsgemeinschaft (DFG, German Research Foundation) - 430621735. Both authors also gratefully acknowledge funding by the Unversity of Bamberg through Fres(c)h grant no. 06999902.
\item Conflict of interest/Competing interests: The authors declare that they have no competing interests.
\item Ethics approval: Our reseach did not involve any human participants, human material, or human data. Our manuscript is not under consideration elsewhere and has not already been published in any form or language. The paper partially builds on and extends results published as ``A Network-Based Explanation of Inequality Perceptions'', published in Social Networks vol. 70, as referenced within the main body of text. 
\item Consent to participate: Not applicable.
\item Consent for publication: Not applicable.
\item Availability of data and materials: Not applicable.
\item Code availability: The code and documentation are available through a GitHub repository at \href{https://github.com/mayerhoffer/Inequality-Perception}{https://github.com/mayerhoffer/Inequality-Perception}.
\item The order of authors was determined by dice roll. Individual authors' contributions:
\begin{itemize}
	\item Schulz: Conceptualisation, Validation, Formal analysis, Writing, Visualisation, Funding acquisition
	\item Mayerhoffer: Conceptualisation, Methodology, Software, Validation, Formal analysis, Writing, Funding acquisition
\end{itemize}
\end{itemize}

\bibliography{consumptionbibliography}


\end{document}